\documentclass[twocolumn,prl,showpacs,preprintnumbers,amsmath,amssymb]{revtex4}

\usepackage{amsmath,bbm}
\usepackage{graphicx,subfigure}
\usepackage{enumitem}

\newcommand{\bra}[1]{\langle #1|}
\newcommand{\ket}[1]{|#1\rangle}
\newcommand{\braket}[1]{\langle #1 \rangle}

\newcommand{\bbi}{\mathbbm{i}}
\newcommand{\tr}{\text{tr}}
\newcommand{\id}{\mathbbm{1}}

\begin{document}

\bibliographystyle{prsty} 

\title{Quantum benchmarks for the storage or transmission of quantum light
   from minimal resources}
\date{\today}

\author{Hauke H\"aseler and Norbert L\"utkenhaus}
\affiliation{
Institute for Quantum Computing and Department of Physics and Astronomy, University of Waterloo, ON, N2L 3G1, Canada\\
Quantum Information Theory Group, Institut f\"ur Theoretische Physik I, and Max Planck Institute for the Science of Light, University of Erlangen-N\"urnberg, 91058 Erlangen, Germany}

\begin{abstract}
We investigate several recently published benchmark criteria for storage or transmission of continuous-variable quantum information. A comparison reveals that criteria based on a Gaussian distribution of coherent states are most resilient to noise. We then address the issue of experimental resources and derive an equally strong benchmark, solely based on three coherent states and homodyne detection. This benchmark is further simplified in the presence of naturally occurring random phases, which remove the need for active input state modulation.
\end{abstract}

\pacs{03.67.Hk, 42.50.Ex, 42.50.Xa}

\maketitle

In quantum communication, inevitable interactions with noisy environments preclude the faithful transmission of quantum states in a direct manner. Yet, many protocols rely on near-perfect state transmission. While auxiliary methods, such as quantum error correction \cite{nielsen00a}, can compensate for channel imperfections, this is only possible if the quantum channel meets a fundamental requirement: It must outperform all \emph{measure \& prepare} (MP) schemes, i.e., schemes in which input states are measured, the resulting information is transmitted classically, and new states are prepared accordingly. Mathematically, these schemes correspond to entanglement-breaking channels \cite{horodecki03a}. The output of such channels and the corresponding measurement data are fundamentally restricted, and it is the aim of benchmark criteria to establish whether these restrictions are overcome. In practice, this requires an ensemble of input states $\{p_i, \ket{\phi_i}^{in} \}_i$ and the collection of measurement statistics on the respective output states $\rho_i^{out}$. This comprises the \emph{classical data}, and we say that a channel acts in the \emph{quantum domain} if the obtained classical data is not compatible with any MP scheme.

Deriving such benchmarks is especially challenging in the continuous-variable regime, where the set of input states must be restricted to those which are experimentally accessible. This has lead to a large number of proposed benchmark criteria \cite{braunstein00a, hammerer05a, namiki08a, namiki08b, haseler08a, calsamiglia09a, owari08a, adesso08b, haseler09a}, which differ in the choice of input states and in the underlying validation method. We begin our analysis by comparing the benchmarks based on coherent states with respect to their resilience to channel noise and come to the conclusion that these criteria either require too many experimental resources to be implemented, or they are too difficult to surpass in state-of-the-art teleportation or quantum storage experiments. Therefore, the problem of finding useful success criteria for such experiments must still be considered an open one.

In this Rapid Communication, we solve this problem by showing that the generation of three coherent states and homodyne detection of the output light suffice to derive a benchmark of optimal strength. Furthermore, we investigate the role of the phase reference in typical experimental setups, and find that copies of a single coherent state can lead to the same benchmark.

\begin{figure}
	\centering
	\subfigure{\includegraphics[width=0.15\textwidth]{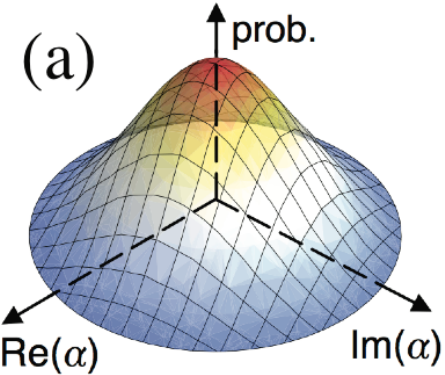}
	\label{fig:in1}
	}
	\subfigure{\includegraphics[width=0.126\textwidth]{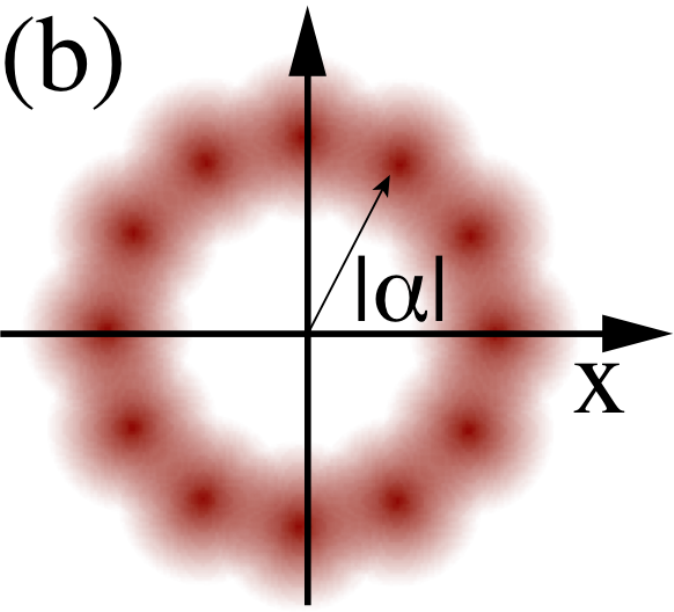}
	\label{fig:in2}
	}
	\subfigure{\includegraphics[width=0.13\textwidth]{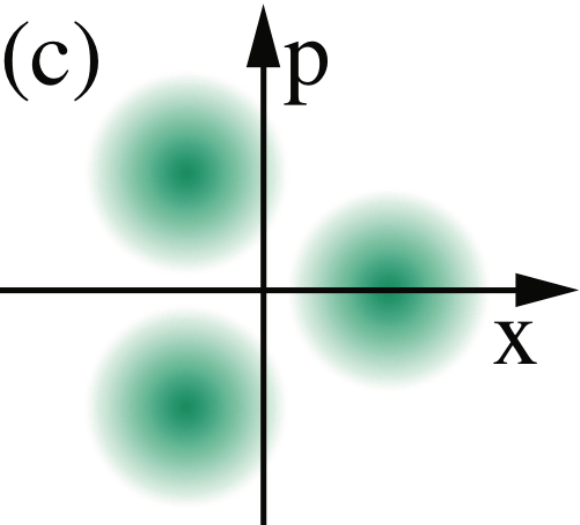}
	\label{fig:in3}
	}
	\caption{(Color online) Different input ensembles of coherent states for testing quantum channels. (a) Gaussian distribution, (b) phase encoding, (c) discrete ensemble.}
	\label{fig:inputs}
\end{figure}


To draw a meaningful comparison between different benchmark criteria, the infinite freedom in the choice of input states and in the nature of the output states must be abridged. For the input states, coherent states are the obvious choice since they are readily generated. To further facilitate a comparison we choose devices which give output states as of a lossy an noisy channel. Let us emphasize here that none of the compared criteria depends on this assumption about the quantum channel, but that it merely serves to form a comprehensive comparison. We choose this model because it reflects the most common imperfections in the transmission or storage of light, namely, photon losses and Gaussian noise. Such a transmission line is conveniently described by a perfect line with an inserted beam splitter and a thermal state entering the beamsplitter's second input port. In this model, the reflectivity $1 - \eta$ of the beamsplitter is a measure of channel loss, and the mean photon number $\bar{n}$ of the thermal state quantifies noise. Within this framework, benchmark criteria differ in the following aspects:

(1) Validation method: The most common method is to maximize the average fidelity \cite{jozsa94b}
	\begin{equation}
		\bar{F}^{max} = \max_{\rho_i^{out}} \sum_i p_i \braket{\psi_i^{in} | \rho_i^{out} | \psi_i^{in}},
	\end{equation}
	for output states $\rho_i^{out}$ resulting from MP channels. Surpassing the resulting benchmark value certifies operation in the quantum domain \cite{braunstein00a}.  An alternative to this is the verification of effective entanglement between the input-state source and the channel output using the expectation value matrix (EVM) \cite{haseler08a}.

(2) Unit gain constraint: If a benchmark is derived by comparing the output states to the input states, the resulting criterion will be unable to detect certain channels in the quantum domain, such as channels of high loss \cite{bowen03a}. A more general approach compares each output state to an optimized target state \cite{namiki08a,namiki08b}.

(3) Input ensembles (Fig.~\ref{fig:inputs}): The number and the distribution of input states can range from infinite sets, such as Gaussian distributions of coherent states \cite{hammerer05a,namiki08b} or phase-modulated input ensembles \cite{calsamiglia09a}, to the smallest possible, binary ensemble \cite{haseler08a,namiki08a}.

\begin{figure}
	\centering
		\includegraphics[width=.48\textwidth]{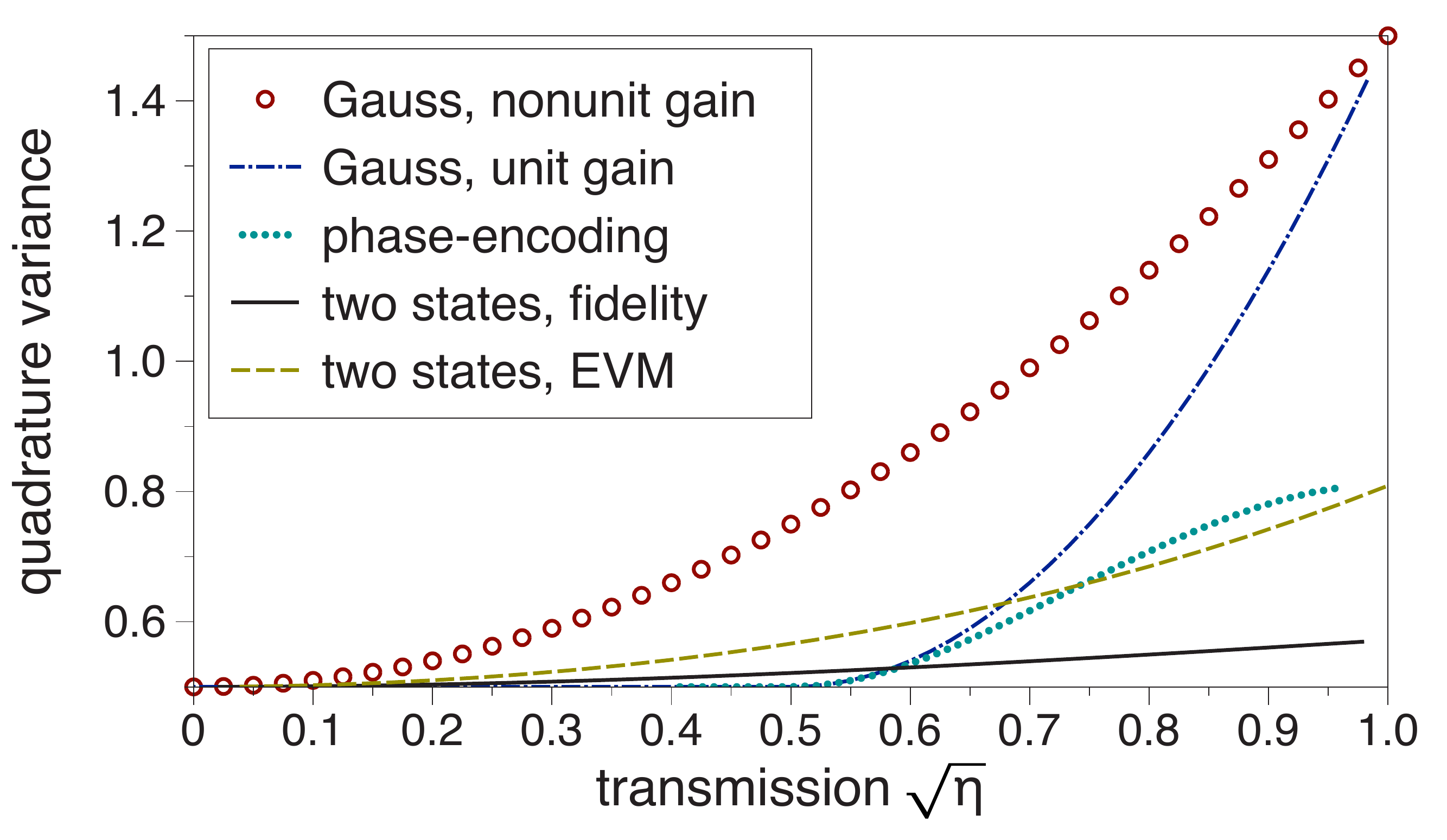}
	\caption{(Color online) Benchmark comparison: channel noise below the curves implies operation in the quantum domain. Circles: Gaussian distribution of input states \cite{namiki08b}. Dashed-dotted: The same with unit-gain constraint \cite{hammerer05a}. Dotted: Ring of coherent states \cite{calsamiglia09a}. Dashed: Two input states, EVM criterion \cite{haseler08a}. Solid: Two input states, fidelity criterion \cite{namiki08a}.}
	\label{fig:comp}
\end{figure}

It is with respect to these aspects that we compare the benchmark criteria presented in Refs.~\cite{hammerer05a,namiki08a,calsamiglia09a,namiki08b,haseler08a}. Each of these criteria has flexibility in the choice of the input ensemble. The maximum classical fidelity for the Gaussian distribution of coherent states, for example, reads $\bar{F}^{max} = (1+\lambda)/(1 + \lambda + \eta)$, where $\lambda$ denotes the inverse width of the input state distribution \cite{namiki08b}. Similarly, benchmarks with phase-modulated coherent input states depend on the chosen amplitude $|\alpha|$. A comparison is then drawn as follows: For each channel loss value, we optimize over the parameter describing the input ensemble to achieve the maximum tolerance to channel noise. Figure \ref{fig:comp} displays the results, where the channel noise is expressed in terms of variances of the quadrature operators $\hat{x} = (\hat{a}^\dagger + \hat{a})/\sqrt{2}$ and $\hat{p} = \bbi (\hat{a}^\dagger - \hat{a})/\sqrt{2}$, rather than in terms of the diverging parameter $\bar{n}$.

A number of qualitative differences between the curves are clearly noticeable. The criteria restricted to unit gain channels, namely those derived by Hammerer \cite{hammerer05a} and Calsamiglia \cite{calsamiglia09a}, are distinct because they can only detect quantum channels for losses below $\sqrt{\eta} = 1/2$. Furthermore, the fidelity-based benchmarks are stronger for larger input ensembles, which has an intuitive explanation: Measuring the average fidelity corresponds to projections onto all input or target states. Hence, larger input ensembles imply more extensive measurements which extract more information on the output states. Lastly, we can draw a comparison between different validation methods. This is only directly possible for the binary input ensemble, where Fig.~(\ref{fig:comp}) shows the verification of effective entanglement \cite{haseler08a} to be significantly stronger than the fidelity-based benchmark \cite{namiki08a}. This may however be a consequence of the quadrature measurements extracting more information on the output states than projections onto two target states.

An unbiased comparison between fidelity-based and entanglement-based validation methods can be drawn for the Gaussian distribution of coherent states, by deriving a criterion which is based on effective entanglement. Generating this ensemble can be thought of as producing a two-mode squeezed state $\ket{r}_{AB} = {\rm sinh}(r) \sum_{n=0}^{\infty} (-{\rm tanh}(r))^n \ket{n}_A \ket{n}_B$, and performing heterodyne detection on system $A$. If the action of the channel on system $B$ does not destroy the initial effective entanglement, operation in the quantum domain is verified. A two-mode squeezed state is represented by a covariance matrix
\begin{equation}
	\frac{1}{2}
	\begin{pmatrix}
		\gamma_A & C \\
		C^T & \gamma_B
	\end{pmatrix},
\end{equation}
with $\gamma_A = \gamma_B = {\rm diag}[\cosh(2r), \cosh(2r)]$ and $C = {\rm diag}[\sinh(2r), -\sinh(2r)]$. The above parameterization of measurement data gives the following output covariance matrix:
\begin{equation}\label{eq:cov}
	\frac{1}{2}
	\begin{pmatrix}
		\gamma_A & \sqrt{\eta} C \\
		\sqrt{\eta} C^T & \eta \gamma_B + (1-\eta) D
	\end{pmatrix},
\end{equation}
with $D = {\rm diag}[1/2 + \bar{n}, 1/2 + \bar{n}]$. The covariance matrix is entangled if its partial transpose is unphysical \cite{simon00a}, which is the case for
\begin{equation}\label{eq:covnoise}
	\bar{n} \le \eta/(1 - \eta).
\end{equation}	
This is the optimal amount of noise tolerance, since it coincides with the particular MP channel described in Ref.~\cite{braunstein00a}. It also coincides with the results obtained by Namiki \emph{et al.}~\cite{namiki08b} by maximizing the classical average fidelity (circles in Fig.~\ref{fig:comp}), but there is one crucial difference: Equation (\ref{eq:covnoise}) was derived for arbitrary non-zero values of the squeezing parameter $r$, i.e., for any width of the Gaussian distribution, whereas the fidelity-based results are shown for the optimized, flat distribution. Evidently, the average fidelity overly emphasizes the \emph{a priory} probabilities $p_i$ and consequently does not lead to the most general benchmark criterion.

Unfortunately, even the approximate generation of a Gaussian distribution requires a very large number of states with high intensities, which makes it impractical for testing quantum channels. All other benchmarks shown in Fig.~\ref{fig:comp} have too little noise tolerance to validate state-of-the-art experiments (see, e.g.~\cite{lobino09a}).\\

\emph{Three-state benchmark:} We now show that a small number $N \ge 3$ of input states and homodyne detection of the output states can lead to a benchmark criterion of optimal strength. Consider the input states $\ket{\phi_j^{in}} = \ket{\alpha \exp(\bbi 2\pi j/N)}$, $j = \{ 1,\dots, N \}$, being picked with equal probability. Their preparation is equivalent to the generation of an effective bipartite state
\begin{equation}\label{eq:effent}
	\ket{\psi_{AB}} = \frac{1}{\sqrt{N}} \sum_{j=1}^{N} \ket{j}_A \ket{\alpha \exp(\bbi 2\pi j/N)}_B,
\end{equation}
followed by projections of system $A$ onto the orthonormal basis elements $\ket{j}_A$ (see, e.g., \cite{curty04a}). Conditioning on the outcomes $j$ reproduces the correct input states for system $B$. The action of the channel on system $B$ transforms the state in Eq.~(\ref{eq:effent}) into a mixed state $\rho_{AB}^{out}$, and the detection of entanglement in $\rho_{AB}^{out}$ verifies that the channel operates in the quantum domain. The classical information available for the verification are the first and second quadrature moments measured on the conditional output states, and the overlap table $\rho_A = \tr_B (\ket{\psi_{AB}} \bra{\psi_{AB}})$. From this information, an EVM can be constructed in direct analogy to Ref.~\cite{haseler08a}:
\begin{equation*}
	\chi_{AB}^{out} = \frac{1}{N}
	\begin{pmatrix}
		\chi_1 & \chi_{12} & \dots \\
		\chi_{21} & \chi_2 & \dots \\
		\vdots & \vdots & \ddots
	\end{pmatrix},
	\ 
	\chi_j = \langle
	\begin{pmatrix}
		\id & \hat{x} & \hat{p} \\
		\hat{x} & \hat{x}^2 & \hat{x}\hat{p} \\
		\hat{p} & \hat{p}\hat{x} & \hat{p}^2
	\end{pmatrix}
	\rangle_{\rho_j^{out}},
\end{equation*}
where the diagonal blocks $\chi_j$ contain the measurement results on the channel output $\rho_j^{out}$. In the off-diagonal blocks $\chi_{ij}$, only the top-left entries $\tr(\rho_{AB}^{out}\ \ket{i}\bra{j} \otimes \id_B) = \braket{ \phi_j^{in} | \phi_i^{in} }$ are known, while the remaining entries are experimentally inaccessible and must be left as free parameters. Now, the separability of $\rho_{AB}^{out}$ is probed on the level of the EVM, with the help of partial transposition, i.e.,
\begin{equation}
	(\chi_{AB}^{out})^{T_A} \not \ge 0 \ \rightarrow \ \rho_{AB}^{out} \ {\rm is \ entangled}.
\end{equation}
Due to the free parameters in the EVM, this positivity criterion is checked numerically via semidefinite programming \cite{vandenberghe95a}.

\begin{figure}
	\centering
		\includegraphics[width=.48\textwidth]{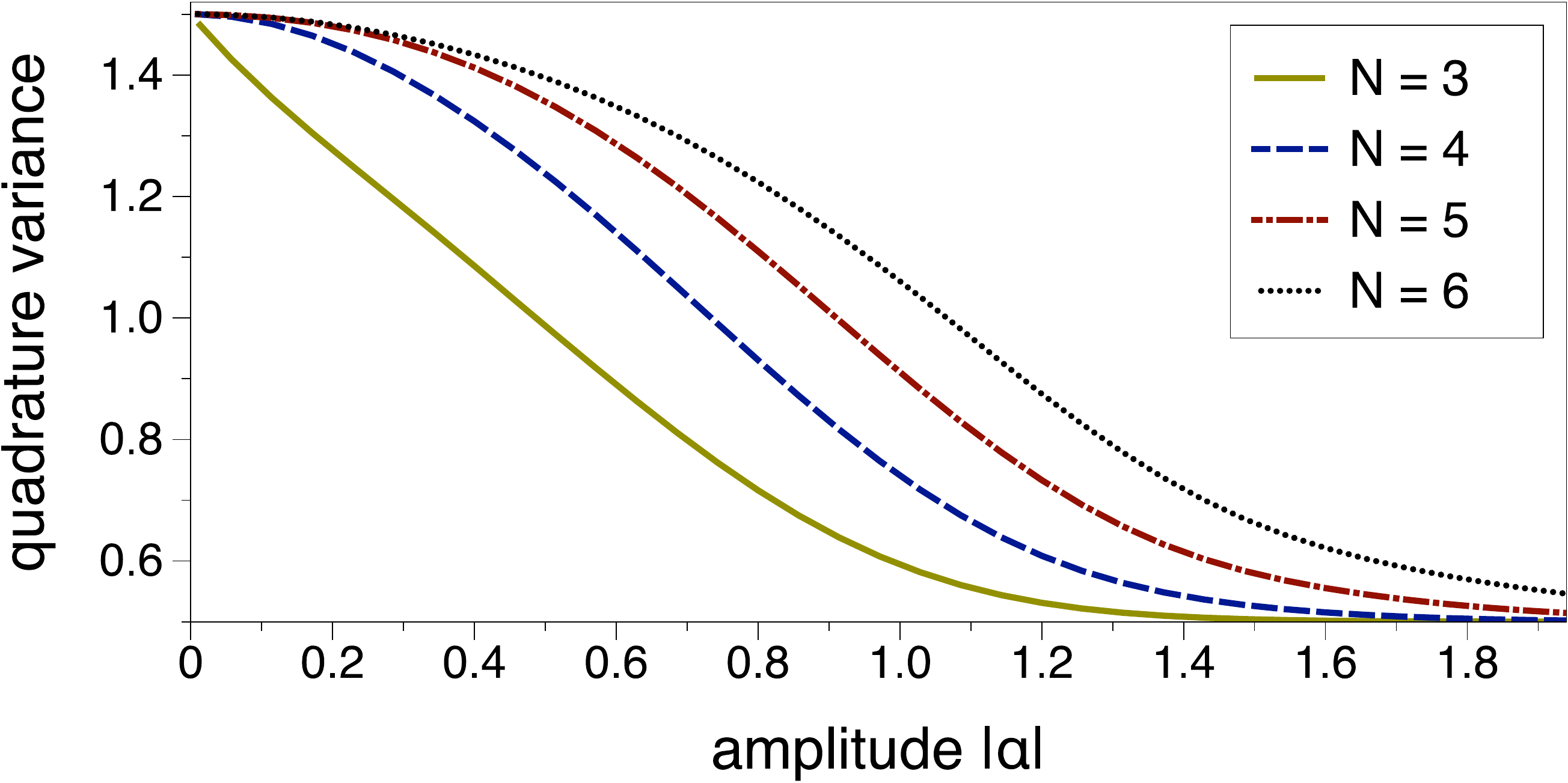}
	\caption{(Color online) Dependence of our benchmarks on the input state intensity for the lossless case. For $\alpha \rightarrow 0$, benchmarks for different $N \ge 3$ coincide.}
	\label{fig:nonzeroalpha}
\end{figure}

This inseparability criterion for $\rho_{AB}^{out}$ allows the identification of a verified quantum domain in terms of the parameters of Fig.~\ref{fig:comp}. Surprisingly, the boundary of this quantum domain coincides, within numerical accuracy, with the non-unit gain benchmark for a Gaussian distribution of input states \cite{namiki08b}, i.e., with the strongest benchmark criterion. This coincidence holds for the optimal working point of the EVM criterion, i.e., $\alpha \rightarrow 0$, and it holds for any $N \ge 3$. In practice, the fact that both measurement statistics and precisions are of finite size will force $\alpha$ away from this optimal working point, the effect of which is shown in Fig.~\ref{fig:nonzeroalpha}. We observe that all curves with $N \ge 3$ coincide for $\alpha \rightarrow 0$, but only for $N \ge 4$ do the curves show a flat slope in that limit. Therefore, the number of test states should be slightly increased when larger coherent amplitudes are employed.

Evidently, testing a channel with three coherent states is far more resilient to loss and noise than using two states. An intuitive understanding of this result may come from state discrimination. The performance of an MP channel naturally depends on the ability to distinguish the input states. Wrong identifications lead to erroneous re-preparations, which will show as broadened observed variances. In discriminating two coherent states $\ket{\alpha}$ and $\ket{-\alpha}$, errors will only occur in the $x$-direction in phase-space, while the $p$-direction remains error free. This can be exploited to limit the variance-broadening induced by the MP channel, as shown in Ref.~\cite{haseler08a}. The situation is different for the above arrangement of three coherent states, where measurement errors will occur along any direction in phase-space, hence leading to more easily detectable MP strategies.

The fact that the preparation of three coherent states can lead to equally strong benchmarks as the Gaussian ensemble is the main result of this Letter. This result ties in with recent findings in continuous-variable quantum key distribution \cite{leverrier09a,sych09suba}. In Ref.~\cite{leverrier09a}, it is argued that a protocol based on the transmission of four coherent states can generate key at a rate at least as high as protocols based on Gaussian modulation, and is in fact able to outperform the latter due to advantages in the classical information processing steps. Our analysis supplies the entanglement verification for such discrete modulated protocols.

Naturally, the above analysis is easily applied to different classes of input states, such as squeezed states, and, following the methods of Ref.~\cite{haseler09a}, mixed states.\\


\emph{Phase Covariance:} We now investigate whether naturally occurring random phases between subsequent signal states can be utilized to encode different input states. Such random phases occur, for example, when using pulsed laser sources.
\begin{figure}
	\centering
	\subfigure[Random phases $\Phi$ of the input states.]{\includegraphics[width=0.45\textwidth]{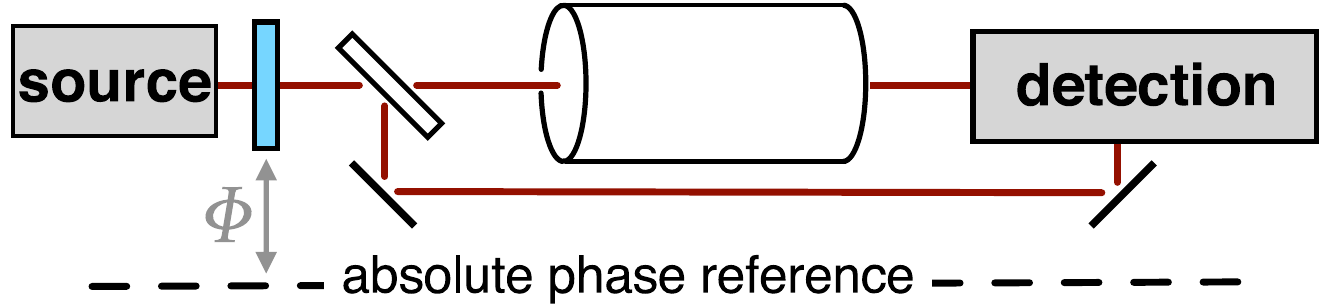}
	\label{fig:subfig1}
	}
	\subfigure[Equivalent phase shifts]{\includegraphics[width=0.45\textwidth]{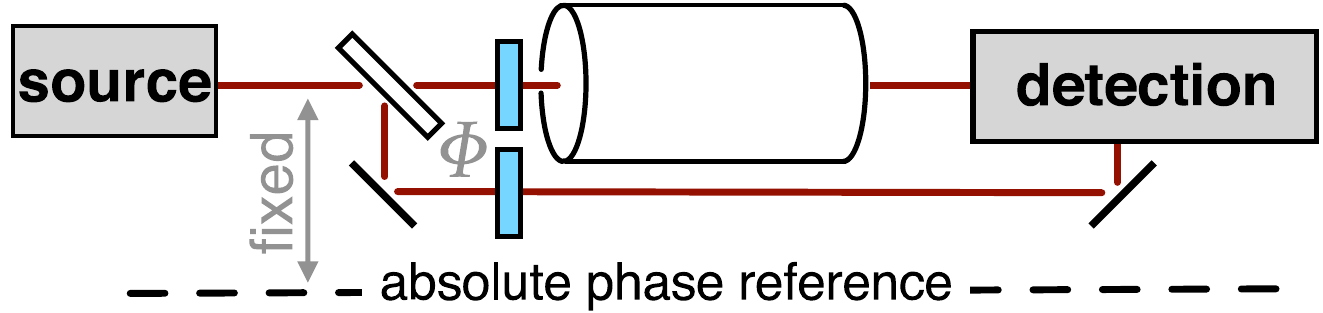}
	\label{fig:subfig2}
	}
	\subfigure[Effectively phase-randomized channel]{\includegraphics[width=0.45\textwidth]{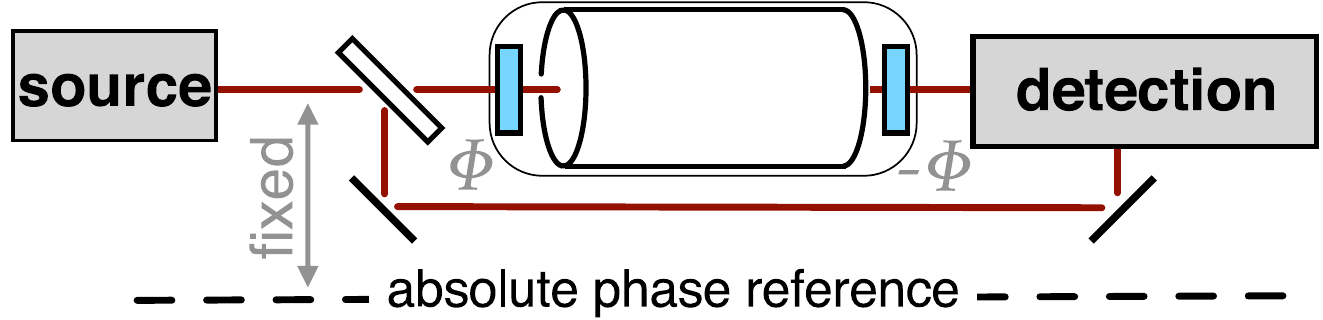}
	\label{fig:subfig3}
	}
	\caption{(Color online) Local oscillator and the theoretical absolute phase reference. Settings \ref{fig:subfig1}, \ref{fig:subfig1} and \ref{fig:subfig1} will lead to the same measurement outcomes.}
	\label{fig:pref}
\end{figure}
Naturally, to speak of the phase of a quantum state implies the existence of a reference frame. In typical setups, both the input states and the phase reference, or \emph{local oscillator}, stem from a master laser, whose output is split on a highly asymmetric beam splitter (see Fig.~\ref{fig:pref}). The encoding of the input states follows after the beamsplitter, by modulating the phase of the input mode with respect to the local oscillator. The question is whether random phases between subsequent pulses emitted by the master laser can replace this encoding step. To speak of the phases of these pulses requires again a reference frame. We will call this frame the \emph{absolute phase reference} (see Fig.~\ref{fig:pref}), which is a theoretically constructed, perfectly bright light beam. The generation of a sequence of phase randomized pulses is equivalent to a train of pulses of equal phase $\Phi = 0$, which are phase randomized after the splitting into signal and local oscillator modes (Fig.~\ref{fig:subfig2}). Now, the phase change $\hat{U}_\Phi$ in the local oscillator mode effectively rotates the axes of the homodyne detection with respect to the absolute phase reference. The same measurement results will ensue if the inverse phase rotation $\hat{U}_{-\Phi}$ is applied to the output of the signal mode (Fig.~\ref{fig:subfig3}). Hence, the random phase shifts between signals are equivalent to wedging the channel between a phase shift and its inverse.
Consequently, the quantum channel is inevitably phase-randomized, or, in other words, the channel is shown to be phase-covariant. This phase-covariance allows us to extract the data necessary for entanglement verification. Suppose an input state $\ket{\alpha}$ leads to a local EVM $\chi_B$, then a phase shifted input state $\hat{U}_\Phi \ket{\alpha}$ will result in the EVM $R(\Phi) \chi_B R(\Phi)^T$, with the rotation matrix
\begin{equation}
	R(\Phi) = 
	\begin{pmatrix}
		1 & 0 & 0 \\
		0 & \cos(\Phi) & -\sin(\Phi) \\
		0 & \sin(\Phi) & \cos(\Phi)
	\end{pmatrix}.
\end{equation}
These local EVMs and the table of overlaps of the different input states suffice to build a bipartite EVM for input ensembles of arbitrary size. Therefore, any $N$ in Fig.~\ref{fig:nonzeroalpha} can be achieved with no active phase-encoding required.


In conclusion, we investigated the strengths and weaknesses of several recently published benchmark criteria for the transmission and storage of quantum information. By modeling typical experimental data, we were able to sort the different benchmarks according to their robustness to noise. Taking into account the required experimental resources, we proposed a verification procedure based on the generation of just three weak coherent states and homodyne detection. Finally, by investigating a typical experimental setup, we found that a single input state setting may suffice to implement a very strong benchmark. We acknowledge funding by \emph{Quantum Works}, the OCE, and the NSERC discovery grant, as well as the European Project QAP.


\begin{thebibliography}{10}

\bibitem{nielsen00a}
M.~A. Nielsen and I.~L. Chuang, {\em Quantum Computation and Quantum
  Information} (Cambridge University Press, Cambridge, 2000).

\bibitem{horodecki03a}
M. Horodecki, P.~W. Shor, and M.~B. Ruskai, Rev. Math. Phys. {\bf 15},  629
  (2003).

\bibitem{braunstein00a}
S.~L. Braunstein, C.~A. Fuchs, and H.~J. Kimble, J. Mod. Opt. {\bf 47},  267
  (2000).

\bibitem{hammerer05a}
K. Hammerer {\it et~al.}, Phys. Rev. Lett. {\bf
  94},  150503  (2005).

\bibitem{namiki08a}
R. Namiki, Phys. Rev. A {\bf 78},  032333  (2008).

\bibitem{namiki08b}
R. Namiki, M. Koashi, and N. Imoto, Phys. Rev. Lett. {\bf 101},  100502
  (2008).

\bibitem{haseler08a}
H. H\"aseler, T. Moroder, and N. L\"utkenhaus, Phys. Rev. A {\bf 77},  032303
  (2008).

\bibitem{calsamiglia09a}
J. Calsamiglia {\it et~al.}, Phys. Rev. A {\bf
  79},  050301  (2009).

\bibitem{owari08a}
M. Owari {\it et~al.}, New J. Phys. {\bf 10},  113014 (20pp)  (2008).

\bibitem{adesso08b}
G. Adesso and G. Chiribella, Phys. Rev. Lett. {\bf 100},  170503  (2008).

\bibitem{haseler09a}
H. H\"{a}seler and N. L\"{u}tkenhaus, Phys. Rev. A {\bf 80},  042304  (2009).

\bibitem{jozsa94b}
R. Jozsa and B. Schumacher, J. Mod. Opt. {\bf 41},  2343  (1994).

\bibitem{bowen03a}
W. Bowen {\it et~al.}, IEEE J. Quant. Electron. {\bf 9},  1519  (2003).

\bibitem{simon00a}
R. Simon, Phys. Rev. Lett. {\bf 84},  2726  (2000).

\bibitem{lobino09a}
M. Lobino, {\it et~al.}, Phys. Rev. Lett. {\bf 102},  203601  (2009).

\bibitem{curty04a}
M. Curty, M. Lewenstein, and N. L\"utkenhaus, Phys. Rev. Lett. {\bf 92},
  217903  (2004).

\bibitem{vandenberghe95a}
L. Vandenberghe and S. Boyd, SIAM Review {\bf 38},  49  (1996).

\bibitem{leverrier09a}
A. Leverrier and P. Grangier, Phys. Rev. Lett. {\bf 102},  180504  (2009).

\bibitem{sych09suba}
D. Sych and G. Leuchs,  Opt. Spectrosc. {\bf 108}, 326 (2010).

\end{thebibliography}
\end{document}